\documentclass[11pt,fleqn]{article}
\usepackage{amsfonts, epsfig, amssymb, amsmath}
\usepackage[normalem]{ulem}
\usepackage{color}
\usepackage{tikz}
\usepackage{pgfplots}
\usepackage{float}

\newcommand{\ol}{\setlength{\itemsep}{0pt.}\begin{enumerate}}
\newcommand{\eol}{\end{enumerate}\setlength{\itemsep}{-\parsep}}
\newcommand{\ignore}[1]{}
\title{An improved bound on $\ell_q$ norms of noisy functions}
\author{Alex Samorodnitsky\thanks{School of Engineering and Computer Science,
The Hebrew University of Jerusalem,
Jerusalem 91904, Israel. Research partially supported by ISF
grant 1724/15.}}

\begin{document}
\date{}
\maketitle


\newtheorem{THEOREM}{Theorem}[section]
\newenvironment{theorem}{\begin{THEOREM} \hspace{-.85em} {\bf :}
}%
                        {\end{THEOREM}}
\newtheorem{LEMMA}[THEOREM]{Lemma}
\newenvironment{lemma}{\begin{LEMMA} \hspace{-.85em} {\bf :} }%
                      {\end{LEMMA}}
\newtheorem{COROLLARY}[THEOREM]{Corollary}
\newenvironment{corollary}{\begin{COROLLARY} \hspace{-.85em} {\bf
:} }%
                          {\end{COROLLARY}}
\newtheorem{PROPOSITION}[THEOREM]{Proposition}
\newenvironment{proposition}{\begin{PROPOSITION} \hspace{-.85em}
{\bf :} }%
                            {\end{PROPOSITION}}
\newtheorem{DEFINITION}[THEOREM]{Definition}
\newenvironment{definition}{\begin{DEFINITION} \hspace{-.85em} {\bf
:} \rm}%
                            {\end{DEFINITION}}
\newtheorem{EXAMPLE}[THEOREM]{Example}
\newenvironment{example}{\begin{EXAMPLE} \hspace{-.85em} {\bf :}
\rm}%
                            {\end{EXAMPLE}}
\newtheorem{CONJECTURE}[THEOREM]{Conjecture}
\newenvironment{conjecture}{\begin{CONJECTURE} \hspace{-.85em}
{\bf :} \rm}%
                            {\end{CONJECTURE}}
\newtheorem{MAINCONJECTURE}[THEOREM]{Main Conjecture}
\newenvironment{mainconjecture}{\begin{MAINCONJECTURE} \hspace{-.85em}
{\bf :} \rm}%
                            {\end{MAINCONJECTURE}}
\newtheorem{PROBLEM}[THEOREM]{Problem}
\newenvironment{problem}{\begin{PROBLEM} \hspace{-.85em} {\bf :}
\rm}%
                            {\end{PROBLEM}}
\newtheorem{QUESTION}[THEOREM]{Question}
\newenvironment{question}{\begin{QUESTION} \hspace{-.85em} {\bf :}
\rm}%
                            {\end{QUESTION}}
\newtheorem{REMARK}[THEOREM]{Remark}
\newenvironment{remark}{\begin{REMARK} \hspace{-.85em} {\bf :}
\rm}%
                            {\end{REMARK}}

\newcommand{\thm}{\begin{theorem}}
\newcommand{\lem}{\begin{lemma}}
\newcommand{\pro}{\begin{proposition}}
\newcommand{\dfn}{\begin{definition}}
\newcommand{\rem}{\begin{remark}}
\newcommand{\xam}{\begin{example}}
\newcommand{\cnj}{\begin{conjecture}}
\newcommand{\mcnj}{\begin{mainconjecture}}
\newcommand{\prb}{\begin{problem}}
\newcommand{\que}{\begin{question}}
\newcommand{\cor}{\begin{corollary}}
\newcommand{\prf}{\noindent{\bf Proof:} }
\newcommand{\ethm}{\end{theorem}}
\newcommand{\elem}{\end{lemma}}
\newcommand{\epro}{\end{proposition}}
\newcommand{\edfn}{\bbox\end{definition}}
\newcommand{\erem}{\bbox\end{remark}}
\newcommand{\exam}{\bbox\end{example}}
\newcommand{\ecnj}{\bbox\end{conjecture}}
\newcommand{\emcnj}{\bbox\end{mainconjecture}}
\newcommand{\eprb}{\bbox\end{problem}}
\newcommand{\eque}{\bbox\end{question}}
\newcommand{\ecor}{\end{corollary}}
\newcommand{\eprf}{\bbox}
\newcommand{\beqn}{\begin{equation}}
\newcommand{\eeqn}{\end{equation}}
\newcommand{\wbox}{\mbox{$\sqcap$\llap{$\sqcup$}}}
\newcommand{\bbox}{\vrule height7pt width4pt depth1pt}
\newcommand{\qed}{\bbox}
\def\sup{^}

\def\H{\{0,1\}^n}

\def\S{S(n,w)}

\def\g{g_{\ast}}
\def\xop{x_{\ast}}
\def\y{y_{\ast}}
\def\z{z_{\ast}}

\def\f{\tilde f}

\def\n{\lfloor \frac n2 \rfloor}

\def \E{\mathop{{}\mathbb E}}
\def \R{\mathbb R}
\def \Z{\mathbb Z}
\def \F{\mathbb F}
\def \T{\mathbb T}

\def \x{\textcolor{red}{x}}
\def \r{\textcolor{red}{r}}
\def \Rc{\textcolor{red}{R}}

\def \noi{{\noindent}}

\def \iff{~~~~\Leftrightarrow~~~~}

\def \queq {\quad = \quad}

\def\<{\left<}
\def\>{\right>}
\def \({\left(}
\def \){\right)}

\def \e{\epsilon}
\def \l{\lambda}

\def \ve{\vec{\epsilon}}
\def \vl{\vec{\lambda}}

\def\Tp{Tchebyshef polynomial}
\def\Tps{TchebysDeto be the maximafine $A(n,d)$ l size of a code with distance $d$hef polynomials}
\newcommand{\rarrow}{\rightarrow}

\newcommand{\larrow}{\leftarrow}

\overfullrule=0pt
\def\setof#1{\lbrace #1 \rbrace}

\begin{abstract}

\noi Let $T_{\e}$, $0 \le \e \le 1/2$, be the noise operator acting on functions on the boolean cube $\H$. Let $f$ be a nonnegative function on $\H$ and let $q \ge 1$. In \cite{S} the $\ell_q$ norm of $T_{\e} f$ was upperbounded by the average $\ell_q$ norm of conditional expectations of $f$, given sets whose elements are chosen at random with probability $\l$, depending on $q$ and on $\e$. In this note we prove this inequality for integer $q \ge 2$ with a better (smaller) parameter $\l$. The new inequality is tight for characteristic functions of subcubes.

\noi As an application, following \cite{SS}, we show that a Reed-Muller code $C$ of rate $R$ decodes errors on $\mathrm{BSC}(p)$ with high probability if
\[
R ~<~ 1 - \log_2\(1 + \sqrt{4p(1-p)}\).
\]
This is a (minor) improvement on the estimate in \cite{SS}.

\end{abstract}

\section{Introduction}

\noi We consider contractive properties of the noise operator acting on functions on the boolean cube $\H$. This is an extensively investigated topic with numerous applications (see e.g., \cite{O'Donnell} for some background). One way to quantify the decrease in the $\ell_q$ norm of a function when this function is acted on by the noise operator was suggested in \cite{S}, where the $\ell_q$ norm of the 'noisy version' of $f$ was upperbounded by the average $\ell_q$ norm of conditional expectations of $f$, given sets whose elements are chosen at random with certain explicit probability $\l$, depending on $q$ and on $\e$. Some applications of this inequality were described in \cite{S, SS}. In this note we prove this inequality for integer $q \ge 2$ with a slightly better (smaller) parameter $\l$, which leads to corresponding improvement in the applications.

\noi We introduce some relevant notions and notation. Given a noise parameter $0 \le \e \le 1/2$, the noise operator $T_{\e}$ acts on functions on the boolean cube as follows: for $f:~\H \rarrow \R$, $T_{\e} f$ at a point $x$ is the expected value of $f$ at $y$, where $y$ is a random binary vector whose $i^{\small{th}}$ coordinate is $x_i$ with probability $1-\e$ and $1 - x_i$ with probability $\e$, independently for different coordinates. Namely, $\(T_{\e} f\)(x) =  \sum_{y \in \H} \e^{|y - x|}  (1-\e)^{n - |y-x|}  f(y)$, where $|\cdot|$ denotes the Hamming distance. We will write $f_{\e}$ for $T_{\e} f$, for brevity.

\noi For $0 \le \l \le 1$, let $T \sim \l$ denote a random subset $T$ of $[n]$ in which each element is chosen independently with probability $\l$. Let $\E(f|T)$ be the conditional expectation of $f$ given $T$. This is a function on $\{0,1\}^n$ defined by $\E(f|T)(x) = \E_{y: y_{|T} = x_{|T}} f(y)$.

\noi We prove the following claim.

\thm
\label{thm:better-RE}
For any integer $q \ge 2$, and for any nonnegative function $f$ on $\H$ holds
\[
\log \|T_{\e} f\|_q ~\le~ \E_{T \sim \l} \log \| \E(f|T)\|_q,
\]
with $\l = \l(q,\e) = 1 + \frac{1}{q-1} \cdot \log_2\(\e^q + (1-\e)^q\)$.

\noi We also have
\[
\log \|T_{\e} f\|_{\infty} ~\le~ \E_{T \sim \l} \log \| \E(f|T)\|_{\infty},
\]
with $\l = \l(\infty,\e) = 1 +  \log_2\(1-\e\)$.

\noi These inequalities are tight if $f$ is a characteristic function of a subcube of $\H$.
\ethm

\noi In \cite{S} this inequality was proved for any real $q > 1$, but with a larger parameter $\l(q,\e)$, given (for $q \ge 2$) by $(1-2\e)^{\frac{q}{(2\ln2) \cdot (q-1)}}$.

\rem

\noi Both the arguments here and in \cite{S} follow well-known proofs for the hypercontractive properties of the noise operator on the boolean cube. In \cite{S} we followed the argument of \cite{Gross}, viewing both sides of the inequality as functions of $\e$, and comparing the derivatives of these functions. In this note we follow the approach of \cite{Beckner}, proving the inequality for the one-dimensional cube, and then extending it to any dimension, using the fact that the boolean cube is a product space. This allows for an improvement in the parameter.
It should be mentioned that the one-dimensional claim turns out to be rather difficult, and we are only able to prove it for integer $q \ge 2$. On the other hand, all the applications which we mention here (and in \cite{S, SS} as well) follow from the special case $q = 2$, which is much easier to prove (see Lemma~\ref{lem:q=2}).
\erem

\noi Theorem~\ref{thm:better-RE} makes it possible to improve the parameters in the results in \cite{S, SS} which use the inequality in \cite{S}. We state some of these results, with the new parameters.

\pro
\label{pro:matroid}
Let $r_C(\cdot)$ be the rank function of the binary matroid on $\{1,...,n\}$ defined by a generating matrix of a linear subspace $C$ of length $\H$. Let $0 \le p \le 1$ and let $t = \log_2(1+p)$. Then
\[
\log_2 \E_{S \sim p} \(2^{|S| - r_C(S)}\) \quad \le \quad \E_{T \sim t} \big(|T| - r_C(T)\big).
\]

\noi This inequality holds with equality if $C$ is a subcube.
\epro

\pro
\label{pro:better doubly transitive}

\noi Let $C$ be a doubly transitive binary linear code of rate $R$. Let $\(a_0,...,a_n\)$ be the weight distribution of $C$. For $0 \le i \le n$, let $i^{\ast} = \min\{i,n-i\}$.

\begin{itemize}

\item For all $0 \le i \le n$ holds
\[
a_i ~\le~ 2^{o(n)} \cdot \(\frac{1}{2^{1-R}-1}\)^{i^{\ast}}.
\]

\item For all $0 \le i \le n$ holds
\[
a_i ~\le~ 2^{o(n)} \cdot \left\{\begin{array}{ccc} \frac{|C|}{\(2 - 2^R\)^{i^{\ast}} \(2^R\)^{n-i^{\ast}}} & 0 \le i^{\ast} \le \(1 - 2^{R-1}\) \cdot n \\ \frac{{n \choose {i^\ast}} \cdot |C|}{2^n} & \mathrm{otherwise} \end{array} \right.
\]
\end{itemize}

\epro

\pro
\label{pro:RM_BSC}

\noi Let $C$ be a binary Reed-Muller code of positive rate $0 < R< 1$. Then $C$ decodes errors on $\mathrm{BSC}(p)$ with high probability (more precisely, a family of such codes $\{C_n\}_n$ with $\limsup_n R\(C_n\) \le R$, attains vanishing error probability on $\mathrm{BSC}(p)$ as $n \rarrow \infty$) if
\[
R ~<~ 1 - \log_2\(1 + \sqrt{4p(1-p)}\).
\]
\epro

\noi This paper is organized as follows. We prove Theorem~\ref{thm:better-RE} in Section~\ref{sec:prf-main}. Propositions~\ref{pro:matroid}-\ref{pro:RM_BSC} do not require new proofs since their claims are obtained by substituting the new value of $\l$ from Theorem~\ref{thm:better-RE} in the corresponding claims in \cite{S,SS}. Note that Proposition~\ref{pro:matroid} corresponds to Lemma~1.8 in \cite{S}, and Propositions~\ref{pro:better doubly transitive}-\ref{pro:RM_BSC} to Proposition~1.1 and Corollary~1.4 in \cite{SS}. The only new observation here is that Proposition~\ref{pro:matroid} holds with equality for subcubes, and this follows immediately from the condition for equality in Theorem~\ref{thm:better-RE}.

\section{Proof of Theorem~\ref{thm:better-RE}}
\label{sec:prf-main}

\noi We prove a more general claim. Consider a more general version of the noise operator. For a vector $\ve = \(\e_1,...,\e_n\)$, with $0 \le e_1,...,\e_n \le \frac12$, the operator $T_{\ve}$ acts on functions on the boolean cube as follows: for $f:~\H \rarrow \R$, $T_{\ve} f$ at a point $x$ is the expected value of $f$ at $y$, where $y$ is a random binary vector whose $i^{\small{th}}$ coordinate is $x_i$ with probability $1-\e_i$ and $1 - x_i$ with probability $\e_i$, independently for different coordinates.

\thm
\label{thm:better-RE-general}
For any integer $q \ge 2$, and for any nonnegative function $f$ on $\H$ holds
\[
\log \|T_{\ve} f\|_q ~\le~ \E_{T \sim \vl} \log \| \E(f|T)\|_q,
\]
with $\vl = \(\l_1,...,\l_n\)$, where $\l_i = \l_i(q,\e) = 1 + \frac{1}{q-1} \cdot \log_2\(\e_i^q + \(1-\e_i\)^q\)$.

We also have
\[
\log \|T_{\ve} f\|_{\infty} ~\le~ \E_{T \sim \vl} \log \| \E(f|T)\|_{\infty},
\]
with $\l_i = \l_i(\infty,\e) = 1 +  \log_2\(1-\e_i\)$.

\noi These inequalities are tight if $f$ is a characteristic function of a subcube of $\H$.

\ethm

\noi We start with the one-dimensional case.

\pro
\label{pro:1-dim}
Let $q \ge 2$ be an even integer. Let $f$ be a nonnegative function on $\{0,1\}$. Then for any $0 \le \e \le \frac 12$ holds
\beqn
\label{ineq:1-dim}
\|f_{\e}\|_q ~\le~ \|f\|_1^{1-\l} \cdot \|f\|_q^{\l},
\eeqn
where $\l = \l(q,\e) = 1 + \frac{1}{q-1} \cdot \log_2\(\e^q + (1-\e)^q\)$.

We also have
\[
\|f_{\e}\|_{\infty} ~\le~ \|f\|_1^{1-\l} \cdot \|f\|_{\infty}^{\l},
\]
with $\l = \l(\infty,\e) = 1 +  \log_2\(1-\e\)$.

\noi These inequalities are tight if $f$ is a characteristic function.
\epro

\noi We will prove this claim in Section~\ref{sec:1-dim} below. For now we assume this claim to hold and proceed with the proof of Theorem~\ref{thm:better-RE-general}.

\noi We introduce the following notation. Let $T_{\e_i}$ denote the noise operator which applies noise $\e_i$ on the $i^{th}$ coordinate. Note that for $\ve = \(\e_1,...,\e_n\)$ holds $T_{\ve} = \prod_{i=1}^n T_{\e_i}$. For a subset $K \subseteq [n]$ of indices, let $T_{\ve, K} = \prod_{i \in K} T_{\e_i}$. Now, let $S$ be the set of indices for which $\e_i \not = 0$. Observe that then $T_{\ve} = T_{\ve,S}$.

\noi Assume $n \ge 2$. The proof will be by induction on the cardinality of $S$. Note that $\e_i = 0$ implies $\l_i = 1$. Hence for $|S| = 0$ the claim amounts to $\|f\|_q \le \|f\|_q$, which is trivial. Assume the claim holds for $|S| = s-1$, and consider the case $|S| = s$. Assume, w.l.o.g., that $n \in S$. Then we have
\[
\|T_{\ve} f\|_q ~=~ \|T_{\ve,S} f\|_q ~=~ \(\E_{x=\(x_1...x_n\)} \(T_{\ve,S} f(x)\)^q\)^{\frac 1q} ~=~ \(\E_{x_1...x_{n-1}} ~\E_{x_n}  \(T_{\e_n} T_{\ve,S \setminus n} f\(x_1...x_n\)\)^q\)^{\frac 1q}.
\]

\noi For each $x_1,...,x_{n-1}$, let $g_{x_1,...,x_{n-1}}$ be the restriction of $T_{\ve,S \setminus n} f$ to the $1$-dimensional cube $\Big(\(x_1,...,x_{n-1},0\), \(x_1,...,x_{n-1},1\)\Big)$. Then the last expression is $\(\E_{x_1...x_{n-1}} ~\E_{x_n}  \(T_{\e_n} g_{x_1...x_{n-1}}\(x_n\)\)^q\)^{\frac 1q}$, and we have
\[
\(\E_{x_1...x_{n-1}} ~\E_{x_n}  \(T_{\e_n} g_{x_1...x_{n-1}}\(x_n\)\)^q\)^{\frac 1q} ~\le~ \(\E_{x_1...x_{n-1}} \(\E_{x_n} g_{x_1...x_{n-1}}\(x_n\)\)^{\(1-\l_n\)q} \cdot \(\E_{x_n} g_{x_1...x_{n-1}}\(x_n\)^q\)^{\l_n}\)^{\frac 1q} ~\le
\]
\[
\(\E_{x_1...x_{n-1}} \(\E_{x_n} g_{x_1...x_{n-1}}\(x_n\)\)^q\)^{\frac{1-\l_n}{q}} \cdot \(\E_{x_1...x_{n}} g_{x_1...x_{n-1}}\(x_n\)^q\)^{\frac{\l_n}{q}},
\]
where in the first inequality we applied the $1$-dimensional inequality, and in the second inequality we have used H\"older's inequality.

\noi Consider the two terms above. Recalling that noise operators commute with conditional expectations, and using the induction hypothesis, the first term is
\[
\(\|T_{\ve, S \setminus n} \E\(f ~|~ \{1...n-1\}\)\|_q\)^{1-\l_n} ~\le~ \mathrm{exp}\left\{\(1-\l_n\) \cdot \E_{T \sim \(\l_1...\l_{n-1},1\)} \ln \|\E\(f|T \cap \{1...n-1\}\)\|_q\right\},
\]
and the second term is
\[
\(\|T_{\ve, S \setminus n} f\|_q\)^{\l_n} ~\le~ \mathrm{exp}\left\{\l_n \cdot \E_{T \sim T \sim \(\l_1...\l_{n-1},1\)} \ln \|\E\(f|T\)\|_q\right\}.
\]

\noi Combining both terms, we get
\[
\|T_{\ve} f\|_q ~=~ \|T_{\ve,S} f\|_q ~\le~
\]
\[
\mathrm{exp}\left\{\(1-\l_n\) \cdot \E_{T \sim \vl} \Big(\ln \|\E\(f|T \)\|_q ~|~  n \not \in T \Big) + \l_n \cdot \E_{T \sim \vl} \Big(\ln \|\E\(f|T \)\|_q ~|~  n \in T \Big)  \right\} ~=
\]
\[
\mathrm{exp}\left\{\E_{T \sim \vl} \ln \|\E\(f|T \)\|_q \right\}.
\]

\noi It remains to show that the inequality in the theorem is tight for characteristic functions of subscube. Let $f$ be such a function. We may assume, by homogeneity, that the expectation of $f$ is $1$. Note that $f$ is a product function, that is $f\(x_1,...x_n\) = \prod_{i=1}^n f_i\(x_i\)$, where each $f_i$ is a function on $\{0,1\}$ which is either twice the characteristic function of $0$ or the constant-$1$ function. In particular, by Proposition~\ref{pro:1-dim}, the $1$-dimensional inequality is tight for each $f_i$. Hence on one hand we have
\[
\|T_{\ve} f\|_q ~=~ \|\prod_{i=1}^n T_{\e_i} f_i\|_q ~=~ \prod_{i=1}^n \| T_{\e_i} f_i\|_q ~=~\prod_{i=1}^n \|f_i\|_1^{1-\l_i} \cdot \|f_i\|_q^{\l_i} ~=~ \prod_{i=1}^n \|f_i\|_q^{\l_i},
\]
where in the last step we have used the fact that the expectation of each $f_i$ is $1$. On the other hand, note that for $T \subseteq [n]$ we have
$E\(f|T\) = \prod_{i \in T} f_i$. Hence
\[
\mathrm{exp}\left\{\E_{T \sim \vl} \ln \|\E\(f|T \)\|_q \right\} ~=~ \prod_{T \subseteq [n]} \|E\(f|T\)\|_q^{\prod_{i \in T} \l_i \prod_{j \not \in T} \(1 - \l_j\)} ~=~ 
\]
\[
\prod_{T \subseteq [n]} \(\prod_{k \in T} \|f_k\|_q\)^{\prod_{i \in T} \l_i \prod_{j \not \in T} \(1 - \l_j\)} ~=~ \prod_{k=1}^n \|f_k\|_q^{\mathrm{Pr}_{T \sim \vl}\left\{k \in T\right\}} ~=~ \prod_{k=1}^n \|f_k\|_q^{\l_k}.
\]

\eprf

\subsection{Proof of Proposition~\ref{pro:1-dim}}
\label{sec:1-dim}

\noi First note that it suffices to prove the claim for finite values of $q$, since the claim for $\|\cdot\|_{\infty}$ follows by taking $q$ to infinity. Note also that the inequalities in the proposition are easily seen to be tight for characteristic functions. In fact, they are trivially true for the constant function, and are easy to verify for a characteristic function of a point.

\noi Fix $q$ and $\e$, which fixes the value of $\l = 1 + \frac{1}{q-1} \cdot \log_2\(\e^q + (1-\e)^q\)$. We may assume, by homogeneity, that $\|f\|_1 = 1$. Under this assumption, we need to show that $\|f_{\e}\|_q \le \|f\|_q^{\l}$, which is equivalent to $\frac{\ln \|f_{\e}\|_q}{\ln  \|f\|_q} \le \l$.

\noi Note that under the assumption $\|f\|_1 = 1$, the function $f$ is determined by its value in $0$, which we denote by $1-x$, $0 \le x \le 1$. Hence, for fixed $q$ and $\e$, the ratio $\frac{\ln \|f_{\e}\|_q}{\ln  \|f\|_q}$ is a univariate function of $x$. It is easy to see that this function equals $\l$ at $x=0$, and we will claim that it indeed attains its maximum in $x = 0$.

\noi It is convenient to introduce the following notation. Let $F(y) = F_q(y) := \ln\(\frac{(1-y)^q + (1+y)^q}{2}\)$. It is easy to see that the function $F$ is increasing for $0 \le y \le 1$. Observe that for $f$ given by values $1-x$ at $0$ and $1+x$ at $1$, we have $\ln \|f\|_q = F(x)$, and $\ln \|f_{\e}\|_q = F((1-2\e) \cdot x)$. So we want to show that for all $0 \le \e \le \frac12$ and for all $0 \le x \le 1$ holds
\[
\frac{F((1-2\e) \cdot x)}{F(x)} ~\le~ \frac{F((1-2\e))}{F(1)}.
\]

\noi Let $G(z) = \ln F\(e^z\)$. Then $G$ is a function on $(-\infty, 0]$, and the inequality above is equivalent to
\[
G(0) + G\Big(\ln(1-2\e) + \ln(x)\Big) ~\le~ G\(\ln(1-2\e)\) + G(\ln(x)),
\]
which will follow if we show that $G$ is concave. From now on we focus on proving the concavity of $G$. We will show that $G'' \le 0$. First, we deal separately with the simple special case $q = 2$, since this suffices for applications.

\lem 
\label{lem:q=2}
The function $G$ is concave if $q=2$.
\elem
\prf
In this case, $F(y) = \ln\(1+y^2\)$, and hence $G(z) = \ln \ln\(1 + e^{2z}\)$. It is easy to see that, up to a positive factor, $G''(z)$ is given by $\ln\(1 + e^{2z}\) - e^{2z}$, which is negative for all $z$. 
\eprf

\noi We continue with the general case. It is easy to see that $G'' \le 0$ is equivalent to
\[
\(yF'' + F'\) \cdot F - y \(F'\)^2 ~\le~ 0.
\]

Writing $H = e^F = \frac{(1-y)^q + (1+y)^q}{2}$, the above is equivalent to, after some rearranging,
\[
\ln(H) ~\le~ \frac{y \(H'\)^2}{yH''H - y\(H'\)^2 + H'H}.
\]

\noi After some (tedious) simplification, we get that
\[
yH''H - y\(H'\)^2 + H'H ~=~ q(q-1)y\(1-y^2\)^{q-2} + \frac{q}{4} \cdot \((1+y)^{2q-2} - (1-y)^{2q-2}\),
\]
and hence that
\[
\frac{y \(H'\)^2}{yH''H - y\(H'\)^2 + H'H} ~=~ \frac{qy\((1+y)^{q-1} - (1-y)^{q-1}\)^2}{4(q-1)y\(1-y^2\)^{q-2} + (1+y)^{2q-2} - (1-y)^{2q-2}}.
\]

\noi So, we need to prove that
\[
\ln(H) ~\le~ \frac{qy\((1+y)^{q-1} - (1-y)^{q-1}\)^2}{4(q-1)y\(1-y^2\)^{q-2} + (1+y)^{2q-2} - (1-y)^{2q-2}}.
\]

\noi Since both sides vanish at $0$, it suffices to prove the inequality for the derivatives, that is, show that
\[
\frac{H'}{H} ~\le~ \frac{d}{dy} ~\frac{qy\((1+y)^{q-1} - (1-y)^{q-1}\)^2}{4(q-1)y\(1-y^2\)^{q-2} + (1+y)^{2q-2} - (1-y)^{2q-2}}.
\]

\noi Let $A(y) = A_q(y) = (1+y)^{q-1}$, and similarly, $B(y) = B_q(y) = (1-y)^{q-1}$. Then, after some simplification, the RHS of the inequality above becomes
\[
\frac{q\(A^2 - B^2\) \cdot \((A-B)^2 + 8(q-1)^2 y^2 \(1-y^2\)^{q-3}\) + q(A-B)^2 \cdot 4(q-1)\(y - 3y^3\)\(1-y^2\)^{q-3}}{\(4(q-1)y\(1-y^2\)^{q-2} + A^2 - B^2\)^2}.
\]

\noi So, we need to verify
\[
\(4(q-1)y\(1-y^2\)^{q-2} + A^2 - B^2\)^2 ~\le~
\]
\[
\frac{H}{H'} \cdot q\(A^2 - B^2\) \cdot \((A-B)^2 + 8(q-1)^2 y^2 \(1-y^2\)^{q-3}\) + q(A-B)^2 \cdot 4(q-1)\(y - 3y^3\)\(1-y^2\)^{q-3}.
\]

\noi Opening up and simplifying, the RHS is
\[
(A+B)^2 \cdot \((A-B)^2 + 8(q-1)^2 y^2 \(1-y^2\)^{q-3}\) +
\]
\[
\(A^2 - B^2\) \cdot 4(q-1)\(y - 3y^3\)\(1-y^2\)^{q-3} +
\]
\[
y \( A^2 - B^2\) \cdot \((A-B)^2 + 8(q-1)^2 y^2 \(1-y^2\)^{q-3}\) +
\]
\[
y (A-B)^2 \cdot 4(q-1)\(y - 3y^3\)\(1-y^2\)^{q-3}.
\]

\noi After some simplification, the inequality becomes
\[
4(q-1)y\(1-y^2\)^{q-3} \cdot \(2\(A^2 - B^2\)\(1-y^2\) + 4(q-1)y\(1-y^2\)^{q-1}\) ~\le
\]
\[
4(q-1)y\(1-y^2\)^{q-3} \cdot \(2(A+B)^2(q-1)y + \(A^2 - B^2\)\(1-3y^2\) + 2\(A^2 - B^2\)(q-1)y^2 + (A-B)^2 \(y-3y^3\)\) +
\]
\[
y\(A^2-B^2\)(A-B)^2.
\]

\noi Clearly, it suffices to prove that
\[
2\(A^2 - B^2\)\(1-y^2\) + 4(q-1)y\(1-y^2\)^{q-1} ~\le~
\]
\[
2(A+B)^2(q-1)y + \(A^2 - B^2\)\(1-3y^2\) + 2\(A^2 - B^2\)(q-1)y^2 + (A-B)^2 \(y-3y^3\).
\]

\noi After rearranging, this is the same as
\[
(2q-4)y\((1+y)^{2q-1} + (1-y)^{2q-1}\)  ~\ge~
\]
\[
\(1-y^2\)\((1-3y)(1+y)^{2q-2} - (1+3y)(1-y)^{2q-2}\) + 2y\(1-3y^2\)\(1-y^2\)^{q-1}.
\]

\noi Next, we change variables. Let $t = \frac{1+y}{1-y}$. Then $t \ge 1$. Substituting, dividing both sides of the above inequality by $\(1-y^2\)^{q-1}$, and multiplying by $(t+1)^3$, the LHS becomes $(4q-8) \(t^2-1\) \(t^q + t^{-q+1}\)$, and the RHS becomes $4\cdot\((4-2t) \cdot t^q - (4t-2) \cdot t^{-q+2} - (t-1)\(t^2 - 4t +1\)\)$. Hence the inequality becomes
\[
(q-2)  \(t^2-1\) \(t^q + t^{-q+1}\) ~\ge~ (4-2t) \cdot t^q - (4t-2) \cdot t^{-q+2} - (t-1)\(t^2 - 4t +1\),
\]
or, after multiplying by $t^{q-1}$,
\[
(q-2)  \(t^2-1\) \(t^{2q-1} + 1\) ~\ge~ (4-2t) \cdot t^{2q-1} - (4t-2) \cdot t - (t-1)\(t^2 - 4t +1\)t^{q-1}.
\]

\noi Let $x = t-1$, then $x \ge 0$. Writing the above inequality in terms of $x$, we get
\beqn
\label{ineq-before-series}
(q-2)x(x+2)\((1+x)^{2q-1} + 1\) - (2-2x) (1+x)^{2q-1} + (4x+2)(1+x) + x \(x^2-2x-2\) (1+x)^{q-1} ~\ge~ 0.
\eeqn

\noi From now on we use the assumption that $q$ is an integer. If this is the case, the LHS is a polynomial of degree $2q+1$. We will show that all the coefficients of this polynomial are nonnegative, which will imply its nonnegativity for $x \ge 0$.

\noi Considering the relevant terms, we have that
\[
\mathrm{coef}_{x^0} ~=~ -2 + 2 ~=~ 0,
\]
\[
\mathrm{coef}_{x^1} ~=~ 4(q-2) - 2(2q-1) + 2 + 6 -2 ~=~ 0,
\]
\[
\mathrm{coef}_{x^2} ~=~ (q-2)(2 + 2(2q-1)) - 2 {{2q-1} \choose 2} + 2(2q-1) + 4 - 2 - 2(q-1) ~=~ 0.
\]

\noi For $k \ge 3$ we have
\[
\mathrm{coef}_{x^k} ~=~ - 2 {{2q-1} \choose k} + (2q-2) {{2q-1} \choose {k-1}} + (q-2) {{2q-1} \choose {k-2}} - 2 {{q-1} \choose {k-1}} - 2{{q-1} \choose {k-2}} + {{q-1} \choose {k-3}} ~=
\]
\[
\frac{(2q-1)(2q-2) \cdots (2q - k +2)}{k!} \cdot \(-(q+2) \cdot k^2 + \(4q^2 + 5q + 2\) \cdot k - 4q(2q+1)\) +
\]
\[
\frac{(q-1)(q-2) \cdots (q - k +3)}{(k-1)!} \cdot \(k^2 + (2q-3) \cdot k - \(2q^2 + 4q - 2\)\).
\]

\noi We claim that
\[
\mathrm{coef}_{x^k} ~\ge~ 0 \quad \mathrm{for} \quad 3 \le k \le 2q+1.
\]

\noi We consider two cases: $3 \le k \le q+2$ and $q+3 \le k \le 2q+1$.

\begin{enumerate}

\item $3 \le k \le q+2$.

\noi Clearly $(2q-1)(2q-2) \cdots (2q - k + 3) \ge (q-1)(q-2) \cdots (q - k +3)$, so it suffices to show
\[
(2q - k +2) \cdot \(-(q+2) \cdot k^2 + \(4q^2 + 5q + 2\) \cdot k - 4q(2q+1)\) ~\ge~ -k \cdot \(k^2 + (2q-3) \cdot k - \(4q^2 + 2q - 2\)\).
\]
and
\[
-(q+2) \cdot k^2 + \(4q^2 + 5q + 2\) \cdot k - 4q(2q+1) ~\ge~ 0.
\]
\noi We start with the first of these inequalities. The difference between the LHS and the RHS is
\[
(q+3) \cdot k^3 - 3\(2q^2 + 3q + 3\) \cdot k^2 + 2\(4q^3 + 12 q^2 + 7q + 3\) \cdot k - 8q(q+1)(2q+1).
\]

\noi We view this as a cubic $P(k)$ in $k$, and want to show that this cubic is nonnegative on $[3,q+2]$. First, we check the endpoints of the interval. We have that $P(3) = 2 \cdot \(4 q^3 -3q^2 - 10q + 9\)$. For $q \ge 2$ this is at least $2 \cdot \(5q^2 - 10 q + 9\) \ge 18$. On the other end, we have that $P(q+2) = q\(3q^3 - q - 2\) > 0$.

\noi Next, we claim that $P$ either always increases on the interval, or first increases and the decreases. Since we have checked both endpoints, this will complete the proof. We have that $P'(k) = 3(q+3) \cdot k^2 - 6\(2q^2 + 3q + 3\) \cdot k + 2\(4q^3 + 12 q^2 + 7q + 3\)$. It siffices to check that $P'(3) > 0$ and that either $P'$ is nonnegative throughout, or that the second root of $P'$ is greater than $q+2$ (which means that $P'$ is first positive and then negative). In fact, $P'(3) = 8 q^3 - 12 q^2 - 13q + 33$. For $q\ge 2$ this is at least $4 q^2 - 13q + 33 > 0$. Next, the discriminant of the quadratic $P'$ is
\[
D(q) ~=~ 36\(2q^2 + 3q + 3\)^2 - 24\(q+3\)\(4q^3 + 12 q^2 + 7q + 3\).
\]

\noi There are two cases. First, $D(q) < 0$, in which case $P'$ is always positive. It is not hard to check that this is the case for $q=2,3,4$. The other case is $D(q) \ge 0$, in which case the second root of $P'$ is given by $\frac{6\(2q^2 + 3q + 3\) + \sqrt{D(q)}}{6(q+3)}$. We claim that this is larger than $q+2$. In fact, we claim that $\frac{2q^2 + 3q + 3}{q+3} > q+2$, which is easily seen to be true for $q \ge 3$.

\noi Next, we verify that
\[
(q+2) \cdot k^2 - \(4q^2 + 5q + 2\) \cdot k + 4q(2q+1) ~\le~ 0
\]

\noi Let $Q(k) = (q+2) k^2 - \(4q^2 + 5q + 2\) \cdot k + 4q(2q+1)$. We need to show that $Q \le 0$ for $3 \le k \le q+2$. With forethought we show $Q$ to be nonpositive on a larger interval, that is for $3 \le k \le 2q+1$.

\lem
\label{lem:Q}
\[
Q ~\le~0, \quad \mbox{for} \quad 3 \le k \le 2q+1.
\]
\elem
\prf
We investigate $Q$ as a quadratic in $k$. Let $A(q) = q+2$, $B(q) = 4q^2 + 5q + 2$, $C(q) = 4q(2q+1)$, and $D(q) = B^2 - 4AC$. Simplifying, we get that $D(q) = 16q^4 + 8q^3 - 39q^2 - 12q + 4$, which is easily seen to be positive for $q \ge 2$, since in this case $16q^4 + 8q^3 \ge 80 q^2$.
The roots of $Q$ are $k_{1,2} = \frac{B \pm \sqrt{D}}{2A}$. We claim that $k_1 \le 3$ and that $k_2 \ge 2q+1$, which will prove what we need.

\noi We start with $\frac{B - \sqrt{D}}{2A} \le 3$, which is the same as $B - 6A \le \sqrt{D}$. It suffices to verify $B^2 - 4AC \ge (B-6A)^2$, which is equivalent to $9A + C \le 3B$. This is the same as $4q^2 + 2q \ge 12$, which is clearly true for $q \ge 2$.

\noi We proceed with $\frac{B + \sqrt{D}}{2A} \ge 2q+1$, which is the same as $\sqrt{D} \ge 2(2q+1)A - B$. It suffices to verify $B^2 - 4AC \ge (2(2q+1)A - B)^2$, which is equivalent to $(2q+1)^2 A + C \le (2q+1) B$. This is the same as $2q^2 \ge 4q$, which is clearly true for $q \ge 2$.
\eprf

\noi This concludes the case $3 \le k \le q+2$.

\item $q+3 \le k \le 2q+1$.

\noi In this case, recalling the assumption that $q$ is integer, we have that $2 {{q-1} \choose {k-1}} + 2{{q-1} \choose {k-2}} - {{q-1} \choose {k-3}} = 0$ and hence, using Lemma~\ref{lem:Q} in the last step, we have
\[
\mathrm{coef}_{x^k} ~=~ - 2 {{2q-1} \choose k} + (2q-2) {{2q-1} \choose {k-1}} + (q-2) {{2q-1} \choose {k-2}} ~=
\]
\[
\frac{(2q-1)(2q-2) \cdots (2q - k +2)}{k!} \cdot (-Q(k)) ~\ge~ 0.
\]

\end{enumerate}

\eprf

\section*{Acknowledgments}

\noi We would like to thank Ori Sberlo for a very helpful discussion.

\end{document}